\documentclass[useAMS,usenatbib]{mn2e}

\usepackage{latexsym}
\usepackage{amssymb}
\usepackage{graphicx}

\usepackage{subfigure}

\newcommand{\msun}{{\rm M_{\odot}}}

\newcommand{\dF}{{^{^*}\!\!F}}
\newcommand{\km}{{\rm\,km}}

\def\fps@figure{bp}%
\def\fps@table{bp}%
\def\fps@plate{bp}%
\def\eps@scaling{1.0}%
\newcommand\epsscale[1]{\gdef\eps@scaling{#1}}%
\newcommand\plotone[1]{%
 \centering
 \leavevmode
 \includegraphics[width={\eps@scaling\columnwidth}]{#1}%
}%
\newcommand\plottwo[2]{%
 \centering
 \leavevmode
 \columnwidth=.45\columnwidth
 \includegraphics[width={\eps@scaling\columnwidth}]{#1}%
 \hfil
 \includegraphics[width={\eps@scaling\columnwidth}]{#2}%
}%
\newcommand\plotfiddle[7]{%
 \centering
 \leavevmode
 \vbox\@to#2{\rule{\z@}{#2}}%
 \includegraphics[%
  scale=#4,
  angle=#3,
  origin=c
 ]{#1}%
}%

\newcommand\apj{\rmfamily{ApJ}}%
\newcommand\apjl{\rmfamily{ApJ}}%
\newcommand\aap{\rmfamily{A\&A}}%
\newcommand\mnras{\rmfamily{MNRAS}}%
\let\la=\lesssim            
\let\ga=\gtrsim
\title[Relativistic Force-Free Electrodynamic Simulations of Neutron
  Star Magnetospheres]{Relativistic Force-Free Electrodynamic Simulations of Neutron
  Star Magnetospheres}

\author[Jonathan C. McKinney]{Jonathan
  C. McKinney\thanks{E-mail:jmckinney@cfa.harvard.edu}\\ Institute for
  Theory and Computation, Harvard-Smithsonian Center for Astrophysics,
  60 Garden Street, MS 51, Cambridge, MA 02138, USA}

\begin{document}
\date{Accepted 2006 January 18. Received 2006 January 17; in original form 2005
  December 8}
\pagerange{\pageref{firstpage}--\pageref{lastpage}} \pubyear{2006}
\maketitle
\label{firstpage}

\begin{abstract}

The luminosity and structure of neutron star magnetospheres are
crucial to our understanding of pulsar and plerion emission.  A
solution found using the force-free approximation would be an
interesting standard with which any model with more physics could be
compared. Prior quasi-analytic force-free solutions may not be
stable, while prior time-dependent magnetohydrodynamic models used
unphysical model parameters. We use a time-dependent relativistic
force-free electrodynamics code with no free parameters to find a
unique stationary solution for the axisymmetric rotating pulsar
magnetosphere in a Minkowski space-time in the case of no surface
currents on the star.  The solution is similar to the force-free
quasi-analytic solution of \citet{cont99} and the numerical
magnetohydrodynamic solution of \citet{kom05}. The magnetosphere
structure and the usefulness of the classical y-point in the general
dissipative regime are discussed.  The pulsar luminosity is found to
be $L \approx 0.99\pm 0.01 \mu^2\Omega_\star^4/c^3$ for dipole
moment $\mu$ and angular frequency $\Omega_\star$.

\end{abstract}




\section{Introduction}\label{introduction}

Neutron star magnetospheres are suspected to have large regions of
space that are nearly force-free \citep{gj69,rs75}. Self-consistent
quasi-analytic solutions exist that describe the force-free (and some
magnetohydrodynamic (MHD)) parts of the environment of such systems
(see, e.g., \citealt{cont99,goodwin04,gru05}). Quasi-analytic methods
cannot test the stability of their solutions and there is little hope
to study the general nonaligned rotator.

Recently, \citet{kom05} used a time-dependent numerical code to
directly integrate the force-free and MHD equations of motion. They
considered their force-free solution unphysical due to uncontrolled
fast reconnection in the current sheet that led to closed field
lines far beyond the light cylinder.  Their MHD version had no such
defect, but they were forced to introduce unphysical evolution
equations and parameters (see their section 4.2) in order to avoid
numerical errors and strong unphysical features.  They found a
stationary MHD solution that may be unique, but they were uncertain
whether their solution depended upon the unphysical model
parameters.

We investigate axisymmetric neutron star magnetospheres by
integrating the force-free equations of motion using a newly
developed general relativistic force-free electrodynamics (GRFFE)
extension \citep{mckinney05d} to a general relativistic
magnetohydrodynamics (GRMHD) code called HARM \citep{gmt03}.  The
original GRMHD code has been successfully used to study GRMHD models
of accretion flows, winds, and jets
\citep{gmt03,gsm04,mg04,mckinney05a,mckinney05b,mckinney05c}.  The
GRFFE formulation is a simple extension of HARM, so all the code
developed for HARM is immediately useful and has identical
convergence properties to the GRMHD version. We have shown that our
GRFFE formulation is at least as accurate as the GRFFE formulation
by \citet{kom02,kom04} \citep{mckinney05d}. Parabolic spatial
interpolation and fourth-order Runge-Kutta temporal integration are
used to improve accuracy. Compared to \citet{kom05}, our force-free
scheme is improved by significantly lowering the reconnection rate
in current sheets by forcing numerically induced velocities into
current sheets to vanish. This eliminates all the difficulties
encountered by \citet{kom05}.

Section \ref{models} discusses the solution of the neutron star
magnetosphere.  Section \ref{conclusions} summarizes the results of
the paper.  The notation follows \citet{mtw73} and the signature of
the metric is $-+++$.  Tensor components are given in a coordinate
basis.  The components of the tensors of interest are given by
$F^{\mu\nu}$ for the Faraday tensor, $\dF^{\mu\nu}$ for the dual of
the Faraday, and $T^{\mu\nu}$ for the stress-energy tensor. The field
angular frequency is $\Omega_F\equiv
F_{tr}/F_{r\phi}=F_{t\theta}/F_{\theta\phi}$.  The magnetic field can
be written as $B^i=\dF^{it}$.  The poloidal magnetospheric structure
is defined by the $\phi$-component of the vector potential
($A_\phi\equiv\Psi$). The current system is defined by the current
density ($\mathbf{J}$) and the polar enclosed current
($B_\phi\equiv\dF_{\phi t}$).  The electromagnetic luminosity is
$L\equiv -2\pi \int_\theta d\theta T^r_t r^2\sin\theta$.  See
\citet{gmt03,mg04,mckinney05d} for details.

\section{Neutron Star Magnetosphere}\label{models}

Here the implementation details and the solution to the neutron star
magnetosphere are discussed.  It is most useful to the community if
a similar model to that chosen by \citet{kom05} is studied and
compared. This will show that one can study pulsar magnetospheres in
the force-free limit using time-dependent numerical models.  Despite
the unphysical parameters introduced by \citet{kom05}, our solutions
are similar.

The only additional note is that we and they {\rm choose} a solution
with no surface currents on the stellar surface by our and their
choice of relaxed boundary conditions. This corresponds to, e.g., a
neutron star with a crust in shear equilibrium (see, e.g.,
\citealt{ruder98}). Other boundary conditions may lead to other
solutions.

\subsection{Initial and Boundary Conditions}

As in \citet{kom05}, the magnetic field is approximated as an
aligned dipolar field and the space-time is approximated as
Minkowski. The poloidal field at $t=0$ (and for all time for $B^r$)
is set to be the perfect dipole with
\begin{equation}
A_{\phi} = \frac{B_{\rm pole}}{2r} R_{\star}^3\sin^2\theta =
\frac{\mu}{r}\sin^2\theta ,
\end{equation}
where the magnetic dipole moment is $\mu = B_{\rm pole} R_{\star}^3/2$
for a polar magnetic field strength $B_{\rm pole}$ and stellar radius
$R_{\star}$.

As in \citet{kom05}, the initial conditions are arbitrarily chosen to
have a velocity of $v^i=0$ and $B^\phi=0$, and the proceeding violent
non-stationary evolution eventually relaxes to a steady-state
solution.  Thus, any solution found must be a stable solution.  As in
\citet{kom05}, the model is evolved for $55$ light cylinder times.

A steady-state solution with no discontinuities or surface currents
on the stellar surface is found by choosing boundary conditions
determined by an analysis of the Grad-Shafranov equation (see, e.g.,
\citealt{bogo97,beskin97}). One is required to specify $2$
constraints and to fix the magnetic field component perpendicular to
the star.  As in \citet{kom05}, $E_\phi\equiv F_{t\phi}=0$,
$\Omega_F=\Omega_{\star}$, and $B^r$ are fixed in time, where
$\Omega_\star$ is the stellar angular velocity.  This assumes the
particle acceleration gap is negligible, which is not generally true
(see, e.g., \citealt{ms94}).  The velocity obeys the frozen-in
conditions (see equation 46 in \citealt{mckinney05d}) in
steady-state and axisymmetry.

\citet{kom05} set the radius of the star to be $R_{\star}=0.1 R_L$,
where $R_L=c/\Omega_F=c/\Omega_{\star}$ is the light cylinder.  For
$R_{\star}=10\km$, this means they chose a spin period of
$\tau\approx 2.1$ms or $\Omega_{\star}\approx 0.0207 c^3/GM$ for a
neutron star with $M=1.44\msun$.  We choose a similar frequency of
$\Omega_{\star}\approx 0.0216 c^3/GM$ such that the light cylinder
is at $R_L=46.3GM/c^3$.

In practice, $B^r$ and $\Omega_F=\Omega_{\star}$ are fixed, while
$B^\theta$ and $B_\phi$ are parabolically extrapolated into the
neutron star surface from the computational domain.  For a
stationary, axisymmetric force-free solution, one can show that the
field geometry completely determines the velocity (see equation 47
in \citealt{mckinney05d}).  For {\it arbitrary} field components
$B^i$ this prescription for $v^i$ {\it always} leads to a time-like
velocity within the light ``cylinders'' \citep{mckinney05d}.

The polar axis boundary condition is such that the perpendicular
fluxes vanish.  The outer boundary condition is obtained by
extrapolating the field into the boundary zones and setting the
velocity as described above, although the outer boundary is chosen
to be far away to avoid reflections back onto the solution.  In the
event that there is a flux from the outer boundary into the
computational grid, that flux is set to zero.

\subsection{Coordinates}

The computation is performed on a set of uniform rectangular
coordinates described by the vector field $x^{(\mu)}$, where each
uniform coordinate is arbitrarily mapped to, e.g., $t,r,\theta,\phi$
in spherical polar coordinates. Apart from the code's ability to
avoid significant numerical reconnection, the ability to choose an
arbitrary $\theta$ grid is crucial to obtain an accurate solution
around the equatorial current sheet.

The radial coordinate is chosen to be
\begin{equation}\label{radius}
r = R_0 + e^{x^{(1)}} ,
\end{equation}
where $R_0$ is chosen to concentrate the grid zones toward the
surface (as $R_0$ is increased from $0$ to $R_{\star}$). An inner
radius of $R_{in}=R_\star$, an outer radius of $R_{out}=2315GM/c^2$
($50$ light cylinders as in \citealt{kom05}), and $R_0=0.903
R_\star$ are used.

The $\theta$ coordinate is chosen to be
\begin{equation}\label{theta}
\theta = \pi x^{(2)} + \frac{1}{2} (1 - h(r)) \sin(2\pi x^{(2)}) ,
\end{equation}
where $x^{(2)}$ labels an arbitrary uniform coordinate and $h(r)$ is
used to concentrate grid zones toward or away from the equator.  The
wide dead zone in inner-radial regions and the current sheet in
outer-radial regions are both resolved using
\begin{equation}\label{hofr}
h(r) = \left(\frac{1}{2}+\frac{1}{\pi}{\rm atan}\left(\frac{r-r_s}{r_0}\right)\right)\left(h_{\rm
  outer}-h_{\rm inner}\right)+ h_{\rm inner} ,
\end{equation}
where $h_{\rm inner} = 1.8$, $h_{\rm outer}=0.05h_{\rm inner}$,
$r_0=5GM/c^2$, and $r_s=15GM/c^2$.

Different resolutions were chosen to test convergence.  In terms of
radial vs. $\theta$ resolution, we used $64\times 64$, $80\times
64$, $80\times 128$, $80\times 256$, $80\times 128$, $160 \times
128$, $160 \times 256$, and finally for the fiducial model we used
$480\times 256$ to reach a similar resolution of \citet{kom05}, who
used $496\times 244$ for their final resolution.  We find that all
quantities have converged to within $1-5\%$ at our highest
resolution, where the exact percentage depends on the quantity.

\subsection{Magnetosphere Solution}

First, the code is checked that the solution reaches a steady-state.
The solution does reach a well-defined steady-state by about $t\sim
1000GM/c^3$, or after about 20 light cylinder light crossing times,
out to about $r\sim 1000GM/c^3$ for all of the domain except within
the equatorial current sheet.  The part of the equatorial current
sheet that connects to the closed zone undergoes slow, small-amplitude
oscillations.  The frequency of these oscillations is proportional to
the reconnection rate allowed by our reconnection model, while the
amplitude is inversely proportional to the reconnection rate. Based
upon the timing of the oscillations, the fiducial model has a
reconnection period of $\sim 10 {\rm ms}$ near the light cylinder. In
the limit of an unlimited reconnection rate, there are no oscillations
and the magnetosphere has a larger closed zone that extends somewhat
beyond the light cylinder, similar to \citet{kom05}.

\begin{figure}
\begin{center}
\includegraphics[width=3.0in,clip]{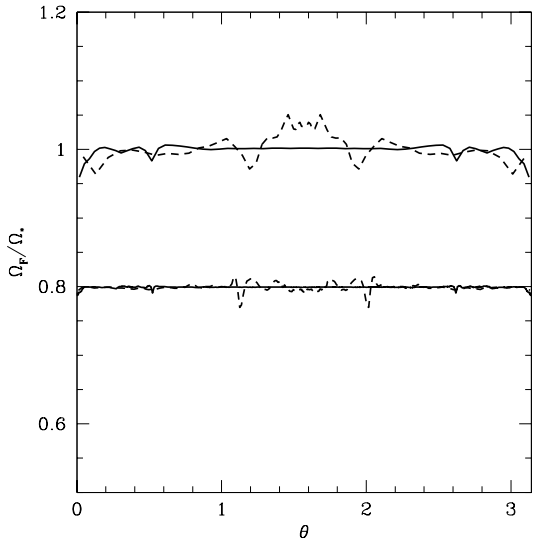}
\end{center}

\caption{Upper two lines show the value of $\Omega_F/\Omega_{\star}$
as a function of $\theta$ for $r=0.2R_L$ (solid line) and $r=0.7R_L$
(dashed line) for $64$ $\theta$ zones and $80$ radial zones. Lower
lines show $\Omega_F/\Omega_{\star}-0.2$ for $256$ $\theta$ zones
and $480$ radial zones. Directly comparable to lower panels in
figure 6 of \citet{kom05}. } \label{nsomegaf2}
\end{figure}

Second, the code is checked for proper integration.  Since the
neutron star has $\Omega_{\star}=\Omega_F$ over the entire surface,
then in axisymmetry and for a stationary flow, $\Omega_F$ should be
the same for all field lines unless force-free (or ideal MHD) is
violated. Notice that strong super-corotation (sub-corotation) would
artificially enhance (diminish) the spindown luminosity and decrease
(increase) the size of the closed zone. Figure~\ref{nsomegaf2} shows
$\Omega_F/\Omega_{\star}$ as a function of $\theta$ for $r=0.2R_L$
and $r=0.7R_L$, the same locations as in \citet{kom05} for their
figure 6 (lower 2 panels).  The plot is for $t=2546.3GM/c^3\approx
55R_L/c$ for a $\theta$ resolution of only $64$ zones and $80$
radial zones (upper lines) and of $256$ $\theta$ zones by $480$
radial zones (lower 2 lines displaced by $0.2$).  \citet{kom05}'s
figure shows that even at a $\theta$ resolution of $244$ that the
peak-to-peak fractional variation is $22\%$ near the equator and
$45\%$ overall (they have problems at the poles).  Our model with
only a $\theta$ resolution of $64$ has a peak-to-peak variation of
only $<10\%$, and there are no significant artifacts at the poles.
Over the entire domain, the spin of the magnetosphere deviates at
worst by $5\%$. At a resolution of $480\times 256$, the variation of
$\Omega_F/\Omega_{\star}$ over the computational grid out to $r\sim
1000GM/c^3$ (large radii still has initial transients) is $<3\%$.

Finally, the region where the stationary solution is actually
force-free is checked, which is tracked by our code
\citep{mckinney05d}.  Force-free is only violated in the very thin
current sheet and only in the sheet for $r>2.8R_L$.  Energy is not
strictly conserved when force-free is violated, but the volume
occupied by this region is negligible and so energy loss is
negligible and does not affect the luminosity as a function of radius
described below.

\begin{figure}
\begin{center}
\includegraphics[width=3.0in,clip]{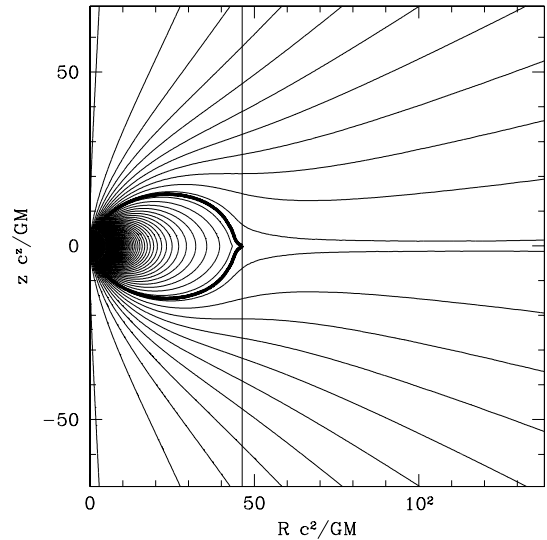}
\end{center}

\caption{Flux function $A_\phi$ in a box with size $3R_L \times
3R_L$ ($R_L$ is light cylinder) as in figure 3 (upper left panel) of
\citet{kom05}. There are 80 contours from $A_\phi=0$ (polar axis) to
$A_\phi=\mu/R_{\star}=9.57\mu\Omega_{\star}/c$ (equator on star). A
single thick solid contour has been added for the closed field line
that touches the theoretical light cylinder. The vertical solid line
is the theoretical light cylinder $R_L$. } \label{nsaphi}
\end{figure}

\begin{figure}
\begin{center}
\includegraphics[width=3.0in,clip]{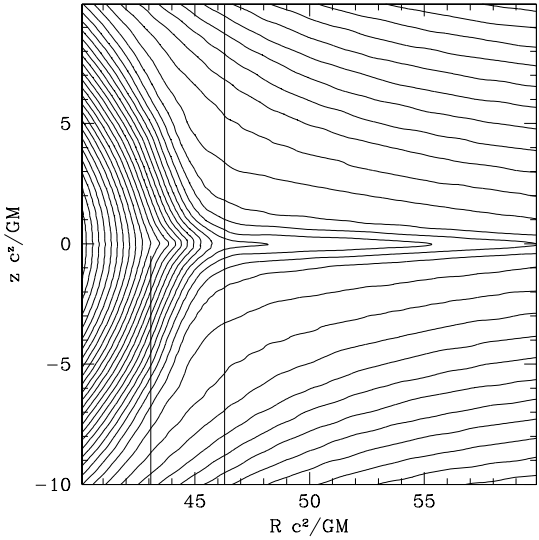}
\end{center}

\caption{Flux function $A_\phi$ for zoomed-in region of
Figure~\ref{nsaphi} around light cylinder. There are 80 contours
from $A_\phi=1.09\mu\Omega_{\star}/c$ to
$A_\phi=1.80\mu\Omega_{\star}/c$. The right vertical line represents
the theoretical light cylinder. The left vertical line points to the
field kink point at the equator. } \label{nsaphizoom}
\end{figure}

Figure~\ref{nsaphi} shows the magnetic flux function ($A_\phi$). The
theoretical light cylinder at $R=R_L$ very closely follows the true
outer light cylinder except in the very thin current sheet, where the
light ``cylinder'' extends slightly radially outward by $\approx 5\%
R_L$. This magnetospheric structure is similar to the force-free
solution of \citet{cont99} (although they have some kinks at the light
cylinder) and the MHD solution of \citet{kom05}.

Figure~\ref{nsaphizoom} shows the structure of the magnetosphere
near the slightly dissipative current sheet. In our solutions and
the solutions of \citet{kom05}, there are always some very thin,
extended closed field loops at the equator.  Using our
low-resistivity model that keeps the current sheet from
reconnecting, the heights of these extended closed field loops are
smaller for increasing resolution, which also more sharply defines
what is qualitatively associated with the so-called y-point. For
general dissipative applications, however, the location of the
y-point is ill-defined and cannot be quantitatively determined.
Notice that while \citet{kom05} state that their closed zone reaches
all the way to the light cylinder, it is clear from their plots and
other statements they make that there is no well-defined closed-open
transition, which is consistent with our results. However, one can
identify at least three quantitative measures, similar to
\citet{kom05}.

First, the first closed field line to develop a kink at the equator
near the light cylinder has an angle of
\begin{equation}
\theta_{\rm kink} \approx 76^\circ \pm 3^\circ
\end{equation}
between the field line and equator at the kink.  This is similar to
the solution found by \citet{gru05}. As they found, we find that this
result is independent of $\Omega_\star$ since the solution is smooth
through the surface. For the model being presented, the intersection
of the kinked field line with the stellar surface is at $\theta_{\rm
kink,\star}\approx 70^\circ$ between a horizontal line and the field
line at the stellar surface.  The kink intersects the equator at
\begin{equation}
r_{\rm kink}\approx 43.08\pm 0.05 GM/c^2=0.931\pm 0.001 R_L,
\end{equation}
slightly within the theoretical light cylinder.  The kink position
oscillates in time by $\approx \pm 0.01R_L$, where the prior error
estimate is smaller since a mean position in time was isolated.  At
the kink, the vector potential has a value
\begin{equation}
{A_\phi}|_{\rm kink} \equiv \Psi_{\rm kink} = 1.35\pm 0.01
\frac{\mu\Omega_\star}{c} ,
\end{equation}
which is quite close to the value of $\Psi_{\rm
open}=1.36\mu\Omega_\star/c$ given by \citet{cont99} and by the value
of $\Psi_{yp}=1.375\pm 0.005\mu\Omega_\star/c$ given by \citet{kom05}.
That the solution kinks before the theoretical light cylinder could be
due to the lack of a perfect dipole due to the light cylinder being
close to the stellar surface, the remaining dissipation in the model,
or could be intrinsic to our stable model.  A comparison of the
numerical dissipative y-point and the theoretical dissipationless
y-point (see, e.g., \citealt{uzdensky03}) is left for future work.

The first kinked field line appears around $r\approx 0.9R_L$ in the
fiducial model of \citet{kom05}, and this agrees with our results.
Since the nature of the dissipation is quite different in each code,
this suggests that our similarly chosen $\Omega_{\star}$ and so the
lack of a perfect dipole, rather than dissipation, is the reason why
the first kinked field line appears inside the theoretical light
cylinder.  Otherwise perhaps it is intrinsic to any stable solution.

Second, in \citet{kom05} they {\it define} the ``y-point'' value of
the vector potential to be at the theoretical light cylinder at
$\theta=\pi/2$, which we find to be
\begin{equation}
{A_\phi}|_{\rm R_L} \equiv \Psi_{\rm light} = 1.27\pm 0.01 \frac{\mu\Omega_\star}{c} ,
\end{equation}
which is quite close to the value of $\Psi_{\rm separatrix}\approx
1.27\mu\Omega_\star/c$ given by \citet{gru05}, similar to $\Psi_{\rm
open}=1.23\mu\Omega_\star/c$ given in \citet{cont05}, similar to
$\Psi_{\rm open}=1.27\mu\Omega_\star/c$ given by \citet{tim05}, and
similar to $\Psi_{\rm open}=1.265\pm 0.005(\mu\Omega_\star/c)$ given
by \citet{kom05}.  Notice that the definition of the y-point value
by \citet{kom05} is their $\Psi_{\rm yp}$, which we defined as
$\Psi_{\rm light}$.  One should compare our $\Psi_{\rm light}$ to
their $\Psi_{\rm yp}$, which are the same measurement.  However,
while they get $1.37$, we get $1.27$ in units of
$c=\Omega_\star=\mu=1$.

Third, as in \citet{kom05}, the vector potential value continues to
drop along the equator until $r>5R_L$ where their value just
oscillates around $\Psi_{\rm open}\approx 1.265\pm
0.005(\mu\Omega_\star/c)$.  We find that such a measurement gives
\begin{equation}
{A_\phi}|_{\rm open} \equiv \Psi_{\rm open} = 1.226\pm 0.005 \frac{\mu\Omega_\star}{c} ,
\end{equation}
which is quite close to the value of $\Psi_{\rm
open}=1.23\mu\Omega_\star/c$ given in \citet{cont05} and slightly
different than the value given by \citet{kom05}.  That the value
given by \citet{kom05} is larger means they have more open flux.
Thus, one expects their luminosity to be larger, which is the case.

Of significant interest is the luminosity coefficient of the dipolar
approximation.  We find that the luminosity is
\begin{equation}
L \approx 0.99\pm 0.01 \frac{\mu^2\Omega_\star^4}{c^3} .
\end{equation}
This result for $L$ agrees with the result by \citet{gru05}, despite
the detailed differences in the location of the kink-point, y-point,
or the amount of open flux vs. closed flux.  Since energy is
explicitly conserved in our numerical method, the energy flux through
each radius follows this same formula for the entire region that has
reached a steady-state and does not violate force-free, unlike in
\citet{kom05} where energy is not explicitly conserved.  Note that for
a lower resolution model, e.g. $160$ radial zones by $128$ $\theta$
zones, that $L\approx 1.10\pm 0.05 $ in units with
$\mu=\Omega_\star=c=1$, which is quite similar to the result by
\citet{kom05}.  Given that our code with $64$ $\theta$ zones generated
less error in $\Omega_F$ than even their model with $244$ $\theta$
zones, then likely their solution with $496\times 244$ radial
vs. $\theta$ zones is simply not as converged as our solution with
$480\times 256$ zones.  However, the agreement between our force-free
model and their ideal MHD model is impressive given the many
unphysical parameters they had to introduce to keep the solution
realistic and the code stable (see their section 4.2).

\begin{figure}
\begin{center}
\includegraphics[width=3.0in,clip]{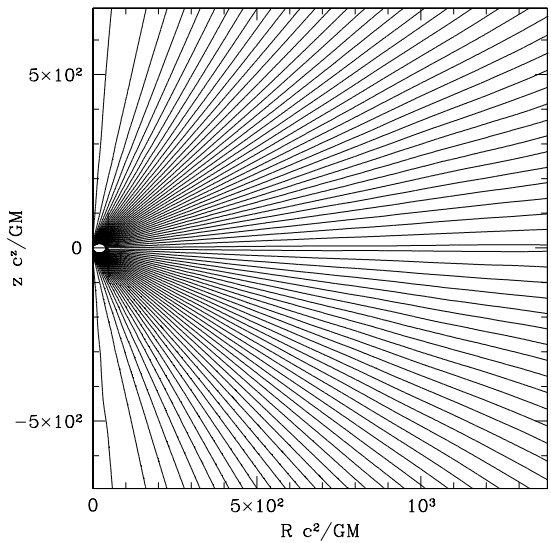}
\end{center}

\caption{Flux function $A_\phi$ out to $30R_L$. There are 40
contours from $0$ to $1.41\mu\Omega_\star/c$. } \label{nsaphilarge}
\end{figure}

Figure~\ref{nsaphilarge} shows the large-scale structure of the
field at the final time of $55R_L/c$ out to $30R_L$.  There are 40
contours from $A_\phi=0$ to $A_\phi=1.41\mu\Omega_\star/c$.  The
field is essentially monopolar.  This is similar to \citet{kom05},
although a reconnection event and plasmoid motion disrupts their
field at $r\sim 23R_L$.

\begin{figure}
\begin{center}
\includegraphics[width=3.0in,clip]{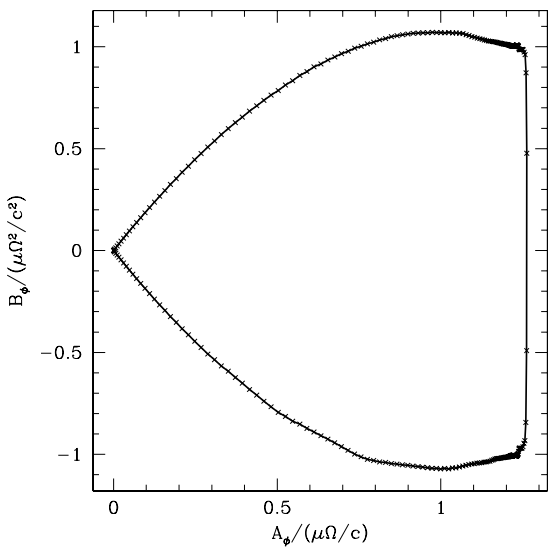}
\end{center}

\caption{Current function $B_\phi$ vs. flux function $\Psi=A_\phi$.
} \label{bphivsaphi}
\end{figure}

A contour plot of the pulsar current function $B_\phi$ shows that
the quantity follows field lines and has a significant current
increase across the separatrix. A plot of the current structure
agrees with the classical model of pulsar current closure and with
\citet{kom05}. To gain quantitative insight, Figure~\ref{bphivsaphi}
shows the pulsar current function vs. the pulsar flux function,
which is a similar plot to figure 4 in \citet{kom05}.  We find that
\begin{equation}
\Psi_{\rm max} = 1.07 \frac{\mu\Omega_\star}{c}~~{\rm at}~~B_\phi =
\pm 1.05 \frac{\mu\Omega_\star^2}{c^2} ,
\end{equation}
at $r=1.1R_L$.  This is similar to the distribution shown by
\citet{cont99} and is also similar to figure 4 in \citet{kom05}.

\section{Conclusions}\label{conclusions}

A stationary force-free solution is found for the neutron star
magnetosphere with a dipolar surface field in Minkowski space-time
with no stellar surface current.  The luminosity follows the
standard dipolar luminosity with a coefficient determined accurately
to be $k=0.99\pm 0.01$.

The difficulties encountered by \citet{kom05} in the force-free
regime were avoided, and the unphysical parameters they introduced
in the MHD regime were avoided. Their MHD solution is similar to our
force-free solution, which suggests that the solution we and they
find is accurate, stable, and may be unique.

\section*{Acknowledgments}

This research was supported by NASA-Astrophysics Theory Program
grant NAG5-10780 and a Harvard CfA Institute for Theory and
Computation fellowship.  I thank Serguei Komissarov and Dmitri
Uzdensky for useful comments.


\appendix




\label{lastpage}

\end{document}